\documentstyle[multicol,aps,prl,epsf,psfig]{revtex}

\begin{document}
\twocolumn[\hsize\textwidth\columnwidth\hsize\csname
 @twocolumnfalse\endcsname

\title  
{
Critical States in a Dissipative Sandpile Model
}
\author 
{
S. S. Manna$^{1,2}$, A. D. Chakrabarti$^2$ and R. Cafiero$^1$
}
\address
{
$^1$P. M. M. H., \'Ecole Sup\'erieure de Physique et Chimie Industrielles, \\
10, rue Vauquelin, 75231 Paris Cedex 05 France \\ 
$^2$Satyendra Nath Bose National Centre for Basic Sciences, \\
Block-JD, Sector-III, Salt Lake, Calcutta 700 091, India \\
}
\maketitle
\begin{abstract}
A directed dissipative sandpile model is studied in the 
two-dimension. Numerical results indicate that the long time steady  
states of this model are critical when grains are dropped only at  
the top or, everywhere. The critical behaviour is mean-field like. 
We discuss the role of infinite avalanches of dissipative models in periodic
systems in determining the critical behaviour of same models in open systems. \\


\end{abstract}

\vskip 0.5 cm
]

  Spontaneous emergence of long range spatio-temporal correlations in 
driven dynamical systems without fine tuning of any control parameter, is the 
concept of Self-Organized Criticality (SOC) \cite{btw,bak,dhar,asm,kada,manna}.
Since its introduction in 1987 \cite{btw}, the precise conditions which
are necessary and sufficient for SOC, are subjected to intense scrutiny. The 
question which attracted much attention is, can one have criticality if there 
is a non-zero rate of bulk dissipation? While some works at the early stages \cite{hwa}
suggested that indeed, the conservation of the transported quantity in the
dynamical rules is a necessity, the later works claimed a negative answer.

  In this paper, we study a directed dissipative sandpile model and
our numerical results indicate that it is critical. We argue,
that a dissipative model may be critical provided the dissipation is not too 
strong and conjecture a criterion to determine the critical behaviour.
  
  In the sandpile model of SOC, sand grains are locally injected and transported on an 
arbitrary lattice. Too many grains cannot be accommodated at any site. 
A site relaxes if the number of grains exceeds certain cut-off and transfers
the grains equally to the neighbouring sites. This transfer process is conservative, 
where no grain is lost or created. At the critical state, cascades of relaxations
follow due to single injection of grains, which are called avalanches. Grains, 
however,  dissipate out of the system through the boundary, otherwise no steady 
state is possible. This is called the Abelian Sandpile Model (ASM) \cite{btw,asm}. A globally driven 
conservative earthquake model is also similarly defined where energy is fed 
uniformly at all sites and transported \cite{bakearth}. This model reproduces
power laws of energy release similar to the Gutenberg-Richter law \cite{guten}.
  
  There are some studies on the dissipative models also. MKK studied a sandpile 
model where a grain can dissipate during a relaxing event, in a probabilistic manner. 
Numerical findings show that the system reaches a sub-critical state with the        
characteristic sizes of the avalanches depending inversely on the probability of dissipation \cite{kiss}.
On the other hand, the dissipative ASM showed criticality with mean-field like critical 
behaviour \cite{bbmodel}. One dimensional version of this model also showed critical behaviour
even with finite driving rate \cite{ali}. The OFC model \cite{ofc} studied the dissipative
earthquake model, where dissipation is controlled by a parameter $\alpha$.
It is claimed that the OFC model is critical for $\alpha_c < \alpha < \alpha_o$,
the conservative value of $\alpha$ being $\alpha_o$, with the critical behaviour depending on $\alpha$
\cite{ofc,socolar,grass,midd}.
The stochastic version of OFC model, however, is shown to loose criticality for
any $\alpha < \alpha_o$ \cite{hakim}.


\begin{center}
\begin{figure}
\epsfxsize=3.5in
\vskip -1.0 cm
\epsffile{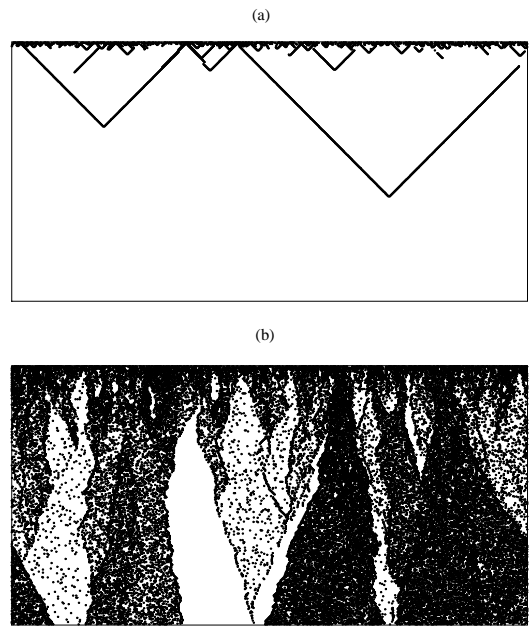}
\vskip -2.9 cm
\caption{
Configurations of DDSM on a lattice of size $L=512$. (a) Avalanches created
by dropping grains randomly only at the top row, orderly place grains along
`V' shaped lines. (b) Grains are dropped at all sites. Different densities
in different regions indicate the age of the big avalnches passed through
that region.
\label{figure1}
}
\end{figure}
\end{center}


  In a conservative model of SOC the grains move a distance of the
order of the system size $L$ when started from the inner most region. This makes the 
average avalanche size  grow as a power of $L$, so that an infinite system has a 
power law distribution of the avalanche sizes. In contrast, in a dissipative 
model, the grains dissipate at any distance within the system. If all grains do 
dissipate within certain cut-off distance, the average avalanche size would not have 
any dependence on $L$ in the large limit. Therefore, for a dissipative 
model to be critical, only a fraction $f(L)$ of grains should dissipate from the bulk and
the rest through the boundary, such that, $f(\infty) = Lt_{L\rightarrow \infty}f(L) \le 1$.
We present examples of both cases in the following.


\begin{center}
\begin{figure}
\epsfxsize=3.5in
\vskip -3.5 cm
\epsffile{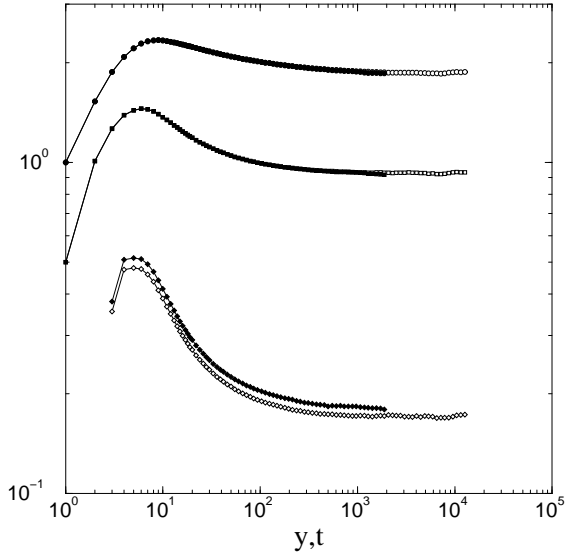}
\vskip -1.0 cm
\caption{
Plots of $P(t)t^{\tau_t-1}$ (circles), $\rho(y)y^{\alpha}$ (squares)
and $\sigma(y)y^{\beta}$ (diamonds) for system sizes $L$ = 2048 (filled symbols)
and 13000 (opaque symbols). Horizontal portions of the curves correspond to
$\tau_t=2.027$, $\alpha=1.012$ and $\beta=1.184$.
\label{figure2}
}
\end{figure}
\end{center}


  On an oriented square lattice with extension $L$, sites are either vacant or occupied
by single grains in the stable state. System is periodic along the $x$ direction and 
the $y$ coordinate increases downward. Grains are dropped randomly. A toppling occurs 
only when the number of grains $h_i > 1$, the site $i$ is then vacated: $h_i \rightarrow 0$. 
The system has a preferrence along the $y$ direction and the down-left and the down-right 
neighbours at the next row gets one grain each: $h_j \rightarrow h_j+1$. In a toppling 
with height 2, grain number is conserved, where as, in a toppling with height 3, one 
grain dissipates from the system. Unlike the Directed Abelian Sandpile model (DASM) 
\cite{dasm}, our model is non-Abelian since sites are vacated in a toppling and we call 
it as the `Directed Dissipative Sandpile Model' (DDSM). 

  In a parallel dynamics all toppling sites reside in a single row on a contiguous toppling 
line (TL). It has a variable length since the fluctuations take place only at the two ends. 
If the TL has a length $\ell$ at time $t$, it would have the length at least $\ell-1$ at time 
$(t+1)$, since all inner $(\ell-1)$ sites will get two grains each and will topple again. If 
either or both neighbours of the end sites are occupied, the TL collects the grains into it, 
and grows in length to $\ell$ or, $\ell+1$. However, if these neighbours are vacant, TL fills 
them and shrinks in length. Therefore, the two ends of the TL move in principle like two 
annihilating random walks starting from the same point. The avalanche terminates when they meet 
and annihilate each other \cite{dasm}. 


\begin{center}
\begin{figure}
\epsfxsize=3.5in
\vskip -4.0 cm
\epsffile{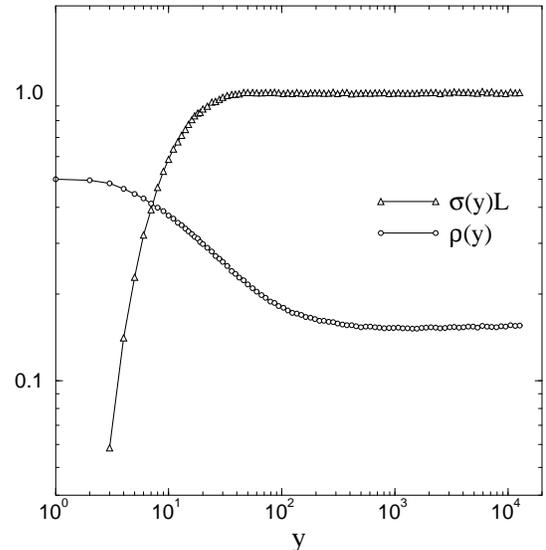}
\vskip -1.0 cm
\caption{
Saturation of the density $\rho(y)$ and the fraction of bulk dissipation
$\sigma(y)$ in a system of size $L=13000$.
\label{figure3}
}
\end{figure}
\end{center}


  Grains are randomly dropped in two ways: In case A, they are dropped only on the top row at 
$y=1$, and in case B, they are dropped everywhere. Grains dissipate through the boundary at $y=L$. 
First, we consider case A and a stable configuration is shown in Fig. 1(a). Grains are marked by black dots, where 
as vacant sites are made blank. Lines of grains in the shape of `V' are mostly observed. This is 
because, due to the bulk dissipation, the density is so low that the TL moves almost in a 
deterministic manner. A `V' is formed by the movement of a TL through a vacant region. In this 
case the TL uniformly shrinks, leaving behind a trail of two converging lines of occupied sites at 
the two ends. However, a TL may also propagate in a `$\Lambda$' between two Vs. In that case it 
uniformly grows in length, deletes two sides of two Vs up to their lowest points and then starts 
shrinking, producing two converging lines which finally make a bigger V. In this way bigger V 
shapes are generated at the expense of smaller Vs, which finally reaches the boundary at the bottom 
and dissipates. Such almost deterministic dynamics makes avalanches of rectangular shape in general, 
but mostly they are squares.

  An avalanche deletes all occupied sites through which it passes. No dissipation occurs 
on the first two rows where the average density is 1/2 as in DASM \cite{dasm}. It then 
decreases with $y$ as a power law: $\rho(y) \sim y^{-\alpha}$. A system of 
size $L = 13000$ is simulated by dropping $2 \times 10^9$ grains. We plot $\rho(y)y^{1.012}$ 
with $y$ on a double logarithmic scale, the curve is horizontal for the large $y$ values,
giving $\alpha = 1.012 \pm 0.030$ (Fig. 2).


\begin{center}
\begin{figure}
\epsfxsize=3.5in
\vskip -4.0 cm
\epsffile{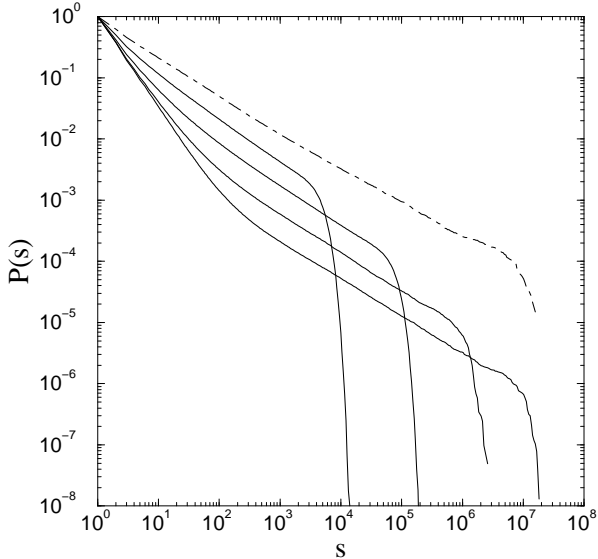}
\vskip -1.0 cm
\caption{
The avalanche size distribution for the case B of DDSM. The dot-dashed curve is
for the high density region. The four curves with solid lines are for avalanches
of the whole system for the system sizes $L$ = 256, 1024, 4096 and 13000 (from top to bottom).
\label{figure4}
}
\end{figure}
\end{center}


  Avalanche size $s$ is the number of sites toppled in an avalanche.
Simulation results indicate that the cumulative probability distribution of $s$ follows
a power law: $P(s) \sim s^{1-\tau_s}$ with $\tau_s = 1.52 \pm 0.03$.
The life-time $t$ of an avalanche is its vertical extension along the preferred
direction and also follows similar power law: $P(t) \sim t^{1-\tau_t}$
with $\tau_t = 2.027 \pm 0.030$ (Fig. 2). The average avalanche size $<s(t)>$ 
varies with life-time $t$ as $<s(t)> \sim t^{\gamma_{st}}$ with
$\gamma_{st} = 2.01 \pm 0.03 $. Since $s, t$ are two measures of the same random avalanche
cluster, they are necessarily dependent, and are related by the scaling
relation: $\gamma_{st} = (\tau_t-1)/(\tau_s-1)$. 

  We explain these results in the following way. It is reasonable to assume that most 
of the avalanches are of rectangular shapes, which implies that $\gamma_{st} = 2$. Now, if 
the TL has a width $w(t')$ at the intermediate time $t'$, then $2w(t')$ grains 
cross that row $y = t'$. The dissipation flux per grain can be divided into `bulk-flux' 
and `boundary-flux'. All grains crossed by the TL, except at its two ends, dissipate.
Therefore, the density and the system size $L$ controll the share between the bulk and the    
boundary fluxes. The constant average boundary-flux through the row at 
$y$ is $<w(y)>y^{1-\tau_t}$ which gives $<w(y)> \sim y^{\tau_t-1}$. But, since average 
avalanche size of life-time ${t}$ is $<s(t)> =\int_0^t w(t')dt' = t^{\tau_t}$, we get 
$\gamma_{st}=\tau_t=2$ and $\tau_s$=3/2. We numerically check the relation 
$<w(y)>=ky$ and find a nice straight line with slope $k=0.312 \pm 0.001$
and the correlation coefficient 0.999.


\begin{center}
\begin{figure}
\epsfxsize=3.0in
\epsffile{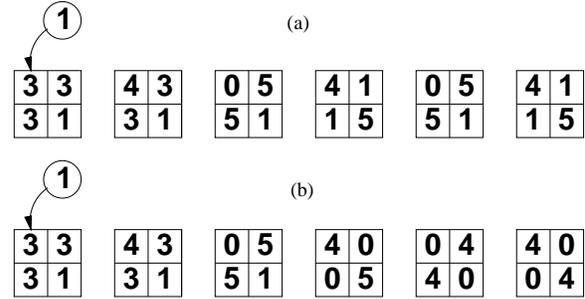}
\vskip 0.5 cm
\caption{
 (a) ASM on a periodic $2 \times 2$ lattice, which leads to a periodic
infinite avalanche (b) Dissipative ASM on a similar periodic lattice also
leads to the periodic infinite avalanche.
\label{figure5}
}
\end{figure}
\end{center}


  Since the density of the system decreases with increasing $y$, we expect that the 
average dissipation also should decrease with increasing $y$. The fraction $\sigma(y)$ 
of total number of grains dissipated in the $y$-th row varies as: 
$\sigma(y) \sim y^{-\beta}$, where $\beta = 1.184 \pm 0.030$ (Fig. 2). 
Therefore the bulk-flux $f(L)$ should vary as: $f(L) = f(\infty) - CL^{-x}$ with
$x=\beta-1$. The exponents $x$ is estimated independently by plotting $f(L)$ vs.
$L^{-x}$ for different $x$ values and the best value obtained is
$x = 0.17 \pm 0.03 $ and $f{(\infty)} = 0.634 \pm 0.010$. 

  Now we consider the case B (Fig. 1(b)). The local density fluctuates widely since an 
avalanche sweeps the region it passes and the local density in this region re-starts 
growing from the scratch. Since the grains are dropped everywhere uniformly and
the bulk dissipation depends on the density, we expect that there should be a saturation
region where the average density is constant. The density $\rho(y)$ decreases from 
1/2 and then saturates to a constant value $0.1543 \pm 0.0010$ around $y_c \approx 100$.
The rate of dissipation $\sigma(y)$, initially increases but finally saturates to the uniform 
dissipation limit: $\sigma(y)=C/L$, with $C=1.01$ (Fig. 3). The bulk-flux $f(L)$ asymptotically
reaches to $f(\infty)$=1 as 1/L. 

  It turned out that the system has two regions. The high density region extends from the
top to $y_c$ and the saturation region from $y_c$ to $L$. We separately collect the 
distribution data for the avalanches 
originated in these two regions. For the high density region, the $\tau_s \approx 1.5$ 
and $\tau_t \approx 2.0$ are obtained as in the case A and $<s(L)> \sim L $ and $<t(L)> \sim \log L$ are 
observed. Linearity in $<w(t)>=k_1t$ is still obeyed with $k_1$=0.1 and 
$\gamma_{st} \approx 2$ is obtained again. However, for the saturation
region, plots of the distribution data showed two regions:
an initial high slope $\tau_s^s \approx 2.5$ for the small $s$ values, followed by a slope 
$\tau_s^l \approx 1.5$ for the large $s$ values. We explain the large value
of $\tau_s^s$ is due to those small avalanches, which grow on an empty
region swept out by a previous large avalanche. When this region reaches the steady state,
the avalanches get the usual exponent $\tau_s^l$ = 1.5 for large $s$ values.
We see that both $<s(L)>$ and $<t(L)>$ have constant values independent of L.
The total distribution, has the behaviour of the saturated regions, since the
avalanches generated in this region have larger weights.

  We now look into the effect of the boundary on dissipative models in more detail.
In a conservative sandpile model with periodic boundary condition, the total mass of the system 
grows up indefinitely. Very soon, an `Infinite Avalanche' starts which never terminates.
For ASM on a periodic square lattice, the same height configuration 
repeats at certain interval, toppling all sites exactly 
once. The period is of the order of $L$ and is dependent on the initial configuration. 
We show such a $2 \times 2$ system in Fig. 5(a). Next we consider the dissipative ASM 
\cite{bbmodel} on the same lattice. After some initial dissipation, 
this model also creates a periodic infinite avalanche which is dissipationless 
(Fig. 5(b)). We now test our DDSM on periodic system, by making the $y$ direction also 
periodic. We observe again 
dissipationless infinite avalanches in both cases A and B. A TL in the form of a ring moves 
indefinitely with uniform speed on the empty torus.

  An infinite avalanche has to be dissipationless after some time, otherwise it will make 
the whole system empty. If a dissipative model in a periodic system has no infinite 
avalanche, it indicates that even the largest avalanche is not as big as the system.
Therefore for the same dissipative model in the open system, the boundary should
have no effect on the avalanche sizes, leading to sub-critical states. We conjecture
that: A dissipative model will not show self-organized criticality if the same model 
on a periodic system has no infinite avalanches. 

  To verify this conjecture, we check 
some examples. The probabilistic dissipation model \cite{kiss}, the stochastic OFC \cite{hakim}
model and the dissipative two states sandpile model \cite{twostate} all are non-critical 
on open boundary systems and do not have infinite avalanches on the periodic systems. 
However, the random creation-dissipation model in \cite{kiss}, the dissipative 
ASM \cite{bbmodel}, the cases of DDSM as described in this paper, lead to SOC states 
with open boundary and also have periodic avalanches on the periodic systems. Finally, 
we check that the deterministic OFC model \cite{ofc} also does not produce any infinite 
avalanche on the periodic system for any $\alpha < 1/4$. Therefore, according to our 
conjecture, deterministic OFC model is not critical, which is against the general belief.
In a recent preprint, it has been claimed that the deterministic OFC model is critical
in the conservative regime only \cite {prado}. 

  We acknowledge D. Dhar with thanks for the critical reading of the manuscript and
for many useful comments. S. S. M. thanks S. Krishnamurthy, S. Roux and S. Zapperi
for usefull discussions. R.C. aknowledges financial support under the European network project
FMRXCT980183.

Electronic address for correspondance: manna@boson.bose.res.in


\end {document}